\documentclass[journal]{IEEEtran}
\usepackage{amsmath,amssymb,amsfonts,bm}
\usepackage{algorithmic}
\usepackage{graphicx}
\usepackage{float}
\usepackage{mdwmath}
\usepackage{booktabs}
\usepackage{tikz}
\usepackage{amssymb}
\usepackage{amsmath}
\usepackage{amsthm}
\usepackage{mathrsfs}

\usepackage{textcomp}
\usepackage{float}
\usepackage{subfig}

\usepackage{soul,color}
\usepackage{cancel}
\usepackage{cite}
\usepackage{cleveref}

\theoremstyle{plain}

\begin{document}

\title{Hermitian Positive Definite Neural Networks for Matrix-CFAR Geometric Signal Detection in Nonhomogeneous Clutter}

\author{
  \IEEEauthorblockN{
  Xiyao Chen\textsuperscript{1}\textsuperscript{*},
  Takumi Sato\textsuperscript{2}\textsuperscript{$\dagger$},
  Linyu Peng\textsuperscript{2}\textsuperscript{$\ddagger$} 
    } \vspace{0.2cm}
    \\
  \IEEEauthorblockA{
    \textsuperscript{1}
Graduate School of Science and Technology, Keio University, Yokohama 223-8522, Japan
  }\\
  \IEEEauthorblockA{
    \textsuperscript{2}
    Department of Mechanical Engineering, Keio University, Yokohama 223-8522, Japan\\
    \textsuperscript{*}Email: chenxiyao@keio.jp\\
 \textsuperscript{$\dagger$}Email: sato8322@keio.jp\\ 
 \textsuperscript{$\ddagger$}Corresponding author. Email: l.peng@mech.keio.ac.jp
  }
}


\maketitle

\begin{abstract}
Matrix information geometry (MIG) detectors exploit the Riemannian geometry of Hermitian positive definite (HPD) matrices to achieve robust object detection in nonhomogeneous clutter. The matrix constant false alarm rate (matrix-CFAR) method is a representative MIG detector that relies on the geometric structures with respect to specific metrics. In most cases, computing representative HPD covariance matrices requires numerical methods, typically involving computationally intensive iterative optimization algorithms. Although explicit formulas are available for certain cases, such as the log-Euclidean (LE) mean, they are generally unavailable for many other matrix means and medians. To address this challenge, we propose integrating the Hermitian positive definite neural network (HPDNet) into the matrix-CFAR framework, replacing conventional numerical optimization procedures with a learned, geometry-preserving mapping that directly estimates representative HPD covariance matrices from data. In the simulations, the framework is evaluated for MIG detectors using both means and medians associated with the affine-invariant Riemannian metric, the LE metric, and the Bures--Wasserstein metric, on both simulated compound-Gaussian clutter and real IPIX sea clutter. In both cases, the HPDNet is trained solely on simulated data to assess its applicability to the study of real data. The results show a significant reduction of computational cost in testing, while maintaining detection performance and robustness.
\end{abstract}

\IEEEpeerreviewmaketitle

\section{Introduction}

The constant false alarm rate (CFAR) method is an adaptive-threshold detection technique widely used in signal detection \cite{PalmerRobertD2008FORS}. It is designed to distinguish targets from clutter by setting a threshold based on the local noise power estimate. Early CFAR variants operate solely on scalar amplitude or power measurements \cite{MaaliA2009ACtf,PacePE1994Faao,BaadecheMohamed2015Paom}. Although they differ in their reference-cell selection strategies, they usually cannot exploit the correlation information contained in multichannel observations, thereby limiting their performance in correlated clutter environments.

Motivated by this drawback, later methods introduced covariance matrix estimation into adaptive detection. Representative covariance-based detectors estimate and exploit the clutter covariance matrix to improve detection performance under homogeneous or nonhomogeneous clutter \cite{ReedIS1974RCRi,HaraY2003Waoa,AbramovichYuriI2004Tado,RobeyFC1992ACam}. Their performance critically depends on the quality of the sample covariance matrix, which requires a sufficient number of homogeneous training samples \cite{RangaswamyM2001Paot}. Beyond these estimation-related issues, these methods are formulated within the Euclidean statistical perspective and do not take advantage of the Riemannian geometry of covariance matrices. These observations have motivated the development of matrix-CFAR detectors, in which covariance matrices are modeled as points on Hermitian positive definite (HPD) manifolds, enabling more faithful characterization of clutter statistics \cite{HuaXiaoqiang2017Gmam}.

To improve detection performance in nonhomogeneous clutter, various matrix information geometry (MIG) detectors have been developed within the matrix-CFAR framework \cite{HuaXiaoqiang2021TDWN,HuaXiaoqiang2022ULDM,OnoYusuke2022Tams,HuaXiaoqiang2023LDfM}. These methods perform detection by measuring geometric distances or divergences between covariance matrices. Since these geometric measures differ substantially in their mathematical formulations, they exhibit different computational complexities and detection characteristics. Thus, the practical deployment of MIG-based detectors is hindered by a trade-off between detection performance and computational efficiency. For example, compared with the log-Euclidean (LE) metric, the affine-invariant Riemannian metric (AIRM) requires computationally intensive matrix operations. This trade-off motivates the development of learning-based approaches that reduce the computational burden of MIG detectors while preserving their detection capability.

 To reduce the computational cost, the concept of a neural network has been introduced to matrix-CFAR MIG detectors. Unlike the calculation in traditional geometric measures such as AIRM, the layer computations in neural networks generally require fewer computational resources. As a result, if the layer structures and their connections are appropriate, neural networks would strike a balance between efficiency and effectiveness. In particular, the symmetric positive definite neural network (SPDNet) was proposed to generalize conventional neural network operations from Euclidean space to Riemannian manifolds \cite{HuangZhiwu2017ARNf}. The Hermitian positive definite neural network (HPDNet) was later developed as a complex-valued extension of SPDNet, so that it can be directly applied to complex covariance matrices which are commonly encountered in signal processing \cite{BrooksDaniel2019AHPD}. These geometric neural networks have recently been explored for covariance matrices in signal detection, showing potential for efficient and robust detection in nonhomogeneous clutter environments \cite{WangRui2025LaBS,WangRui2023DADR}.

In this paper, we apply the HPDNet to the matrix-CFAR framework to reduce the computational cost of geometric signal detection in nonhomogeneous clutter. The main contributions are summarized as follows:
\begin{itemize}
    \item [1.] We propose an HPDNet-based framework for matrix-CFAR MIG detectors for estimating geometric means and medians. Unlike conventional iterative optimization schemes, our framework directly learns the mapping from a set of covariance matrices to their geometric statistics via an HPDNet, eliminating the need for per-test iterative computation and enabling rapid inference once trained.
    \item [2.] We apply the proposed method to both simulated compound-Gaussian clutter and real sea clutter. For the simulated data, training and testing are both conducted on compound-Gaussian clutter. For the real sea clutter data, we train the HPDNet solely on simulated compound-Gaussian clutter and test it directly on the real data, thereby investigating the generalization capability of the learned mapping from simulated clutter environments to practical operational scenarios.
    \item [3.] We compare the proposed HPDNet-based method with conventional geometric estimators across several Riemannian metrics on both simulated and real sea clutter data. On simulated data, the HPDNet improves the detection ability under the AIRM and LE metric while achieving comparable performance under the Bures--Wasserstein  (BW) metric. On real data, the HPDNet achieves superior performance for the AIRM median, LE mean and LE median and maintains comparable performance under the BW metric,
    although a slight degradation is observed for the AIRM mean. In both simulated and real cases, the detection runtime is substantially reduced except for the LE mean. 
\end{itemize}

The remainder of this paper is organized as follows. In Section \ref{sec:CFAR}, we briefly review the matrix-CFAR detection formulation and describe the overall detection scheme. In Section \ref{sec:HPDManifold}, we introduce the geometric structure of HPD manifolds, review the AIRM, LE, and BW metrics, and propose the architecture of HPDNet, including its layer-wise structure and training strategy. In Section \ref{sec:performance}, we report the experimental results obtained from both simulated and real sea clutter data, along with a runtime analysis. Robustness analysis for both the traditional and HPDNet-based methods is provided in Section \ref{sec:robustness}. Finally, in Section \ref{sec:conclusion}, we summarize the main findings of this study and discuss several promising directions for future research.

\section{The Method of Matrix-CFAR} \label{sec:CFAR}
In this section, we describe the objective and basic scheme of the matrix-CFAR method. The aim is to detect whether an object signal is present in the received data returns, under the assumption that the clutter is nonhomogeneous. Let $\mathbf{x} \in \mathbb{C}^{N}$ be the received data, where $N$ denotes the number of observation channels. For simplicity, we assume that the channels are stationary. The detection problem is formulated as a binary hypothesis test between the null hypothesis \(H_0\) (target absent) and the alternative hypothesis \(H_1\) (target present)
\begin{equation}\label{def:h0h1}
    \begin{cases}
        H_0: \begin{cases}
            \text{target is absent},\\
            \mathbf{x} = \mathbf{c},
        \end{cases}\\
        H_1: \begin{cases}
            \text{target is present},\\
            \mathbf{x} = \xi \mathbf{s} + \mathbf{c},
        \end{cases}
    \end{cases}
\end{equation}
where $\mathbf{c}$ represents clutter in $\mathbb{C}^{N}$, $\mathbf{s}$ represents the object signal also in $\mathbb{C}^{N}$, and $\xi$ is a complex-valued coefficient. 

\begin{figure}[htbp]
    \centering
    \includegraphics[width=\linewidth]{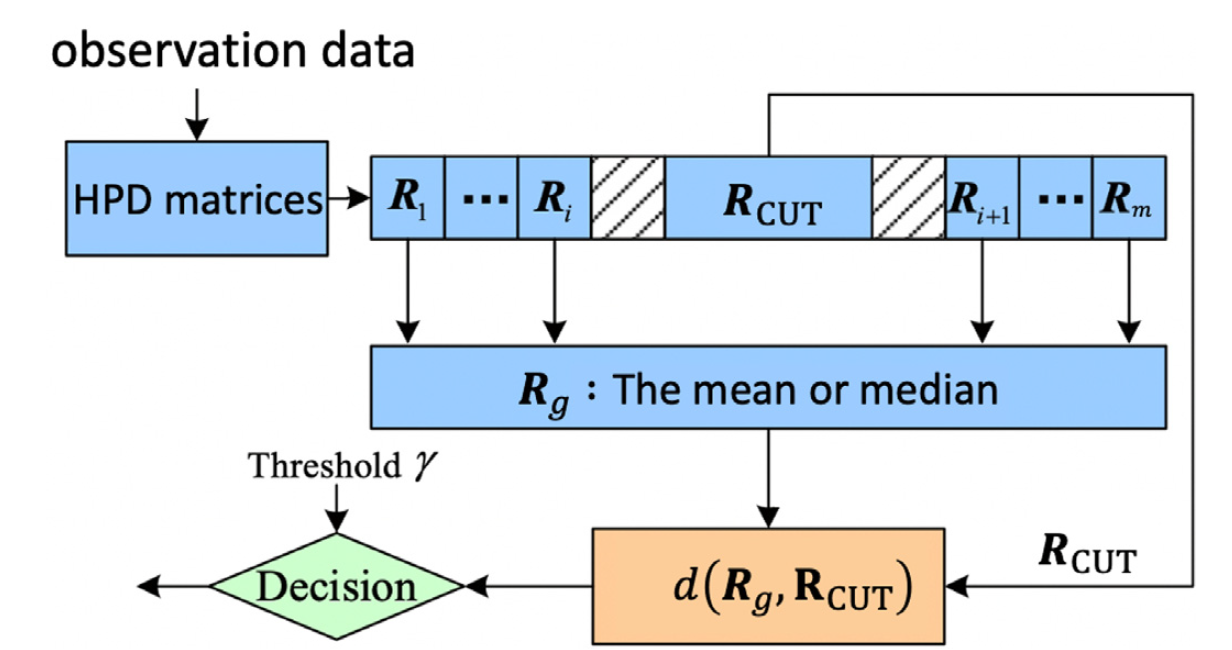}
    \caption{Overview of the matrix-CFAR detection framework.}
    \label{fig:CFAR}
\end{figure}

To exploit the spatial correlation information contained in multichannel observations, matrix-CFAR performs detection on covariance representations rather than raw signal data. A Hermitian Toeplitz covariance matrix is constructed from the estimated autocorrelation sequence as \cite{Kang2019Knowledge}
\begin{equation}\label{def:toeplitz}
\mathbf{R} =
\begin{pmatrix}
\hat r_0  & \cdots & \overline{\hat r}_i  & \cdots & \overline{\hat r}_{N-1} \\
\vdots & \ddots & \ddots & \ddots & \vdots\\
\hat r_i  & \ddots & \hat r_0  & \ddots & \overline{\hat r}_i \\
\vdots & \ddots & \ddots & \ddots & \vdots\\
\hat r_{N-1}  & \cdots & \hat r_i  & \cdots & \hat r_0 
\end{pmatrix},
\end{equation}
where $\overline{z}$ denotes the complex conjugate of $z$, and the autocorrelation coefficient of lag $i$ is estimated as 
\begin{equation}
\hat r_i 
=
\frac{1}{N}
\sum_{n=0}^{N-i-1}
x_{n}\,
\overline{x}_{n+i},
\qquad
0\le i\le N-1.
\end{equation}
The covariance matrices $\{\mathbf{R}_1,\mathbf{R}_2,\ldots,\mathbf{R}_K\}$ estimated from all $K$ units are subsequently processed to generate a representative covariance matrix by
\begin{equation}
    \mathbf{R}_g = \Phi\left(\mathbf{R}_{1}, \mathbf{R}_{2},\ldots, \mathbf{R}_{K}\right),
\end{equation}
where $\Phi(\cdot)$ denotes the geometric mean or median function under a given metric.  Consequently, the detection problem \eqref{def:h0h1} becomes the comparison between covariance matrices,
\begin{equation}
    \begin{cases}
        H_0: \mathbf{R}_g = \mathbf{R}_{\text{CUT}}, \\
        H_1: \mathbf{R}_{g} \neq \mathbf{R}_{\text{CUT}},
    \end{cases}
\end{equation}
where $\mathbf{R}_{\text{CUT}}$ is the covariance matrix estimated from the cell under test (CUT). By adopting a threshold $\gamma$, the decision rule satisfies a prescribed false alarm probability. The threshold $\gamma$ is estimated by Monte Carlo simulations under the null hypothesis $H_0$ and the resulting decision rule is 
\begin{equation}
    d(\mathbf{R}_g,\mathbf{R}_{\text{CUT}}) \overset{H_0}{\underset{H_1}{\lessgtr}} \gamma.
\end{equation}
Here, $d(\cdot,\cdot)$ denotes the distance defined by the adopted Riemannian metric. Fig. \ref{fig:CFAR} illustrates the scheme of the matrix-CFAR method.

\section{Riemannian Structures and the Architecture of HPDNets on HPD Manifolds} \label{sec:HPDManifold}

The proposed framework relies on distances defined on the manifold of HPD matrices. In this section, we will first introduce the geometric tools required throughout the remainder of the paper. In particular, we will review three representative Riemannian metrics, namely, the AIRM,
the LE metric, and the BW metric, along with the algorithms for calculating the corresponding means and medians. We then propose the architecture of the HPDNet, followed by the estimation strategy of the means and medians.

\subsection{Riemannian Metrics}\label{sbusec:HPDmanifoldmetrics}
Since covariance matrices are HPD, they can be naturally regarded as points on the HPD manifold, defined as
\begin{equation}
\begin{aligned}
  \mathscr{P}(N,\mathbb{C}) & = \left\{\mathbf{P} \in GL(N,\mathbb{C}) \mid \mathbf{P} = \mathbf{P}^\mathsf{H}, \right. \\
  & \quad  \left. \mathbf{x}^\mathsf{H} \mathbf{P} \mathbf{x} >0, \forall \mathbf{x} \in \mathbb{C}^N \backslash \{0\}  \right\},   
\end{aligned}
\end{equation}
where $GL(N,\mathbb C)$ denotes the general linear group of all invertible
$N\times N$ complex matrices, and $(\cdot)^{\mathsf H}$ denotes the conjugate transpose. Equipping a tangent space at every point of an HPD manifold provides a local linear structure for geometric computations. The tangent space at a given point $\mathbf{P}$ is defined by
\begin{equation}
    T_{\mathbf{P}}\mathscr{P}(N,\mathbb{C}) \cong \mathcal{H}(N, \mathbb{C}),
\end{equation}
where
\[
\mathcal{H}(N,\mathbb{C})
=
\left\{
\mathbf{X}\in\mathbb{C}^{N\times N}
\,\middle|\,
\mathbf{X}^{\mathsf H}=\mathbf{X}
\right\},
\]
 the vector space of Hermitian matrices, is naturally identified with the local linear space associated with $\mathbf{P}$. A Riemannian metric is defined as an inner product for each tangent space that varies smoothly from one point to another, i.e., 
\begin{equation}
    g({\mathbf{P}}): T_{\mathbf{P}} \mathscr{P} \times T_{\mathbf{P}} \mathscr{P} \to \mathbb{R}.
\end{equation}
Different Riemannian metrics endow the HPD manifold with different notions of distance and geodesic, resulting in different geometric properties and computational costs. 

Among various Riemannian metrics defined on the HPD manifold, the AIRM is one of the most widely used. For any tangent vectors $\mathbf{X}$, $\mathbf{Y} \in T_P\mathscr{P}(N, \mathbb{C})$, the AIRM is defined by \cite{OnoYusuke2024TCoR}
\begin{equation}
    g_{\text{AIRM}}({\mathbf{P}})(\mathbf{X},\mathbf{Y}) = \operatorname{tr}\left(\mathbf{P}^{-1}\mathbf{X}\mathbf{P}^{-1}\mathbf{Y}\right).
\end{equation}
The AIRM induces the geodesic connecting two points $\mathbf{A}, \mathbf{B} \in \mathscr{P}(N, \mathbb{C})$, given by
\begin{equation}
    \boldsymbol{\gamma}(t) = \mathbf{A}^{1/2} \left(\mathbf{A}^{-1/2} \mathbf{B} \mathbf{A}^{-1/2}\right)^t \mathbf{A}^{1/2}, \quad t \in [0,1],
\end{equation}
where $(\cdot)^{1/2}$ denotes the principal HPD matrix square root. The length of this geodesic defines the AIRM distance between $\mathbf{A}$ and $\mathbf{B}$,
\begin{equation}\label{def:AIRMdistance}
    d_{\text{AIRM}}(\mathbf{A},\mathbf{B}) = \|\log(\mathbf{A}^{-1/2}\mathbf{B}\mathbf{A}^{-1/2})\|_F.
\end{equation}
Here, $\|\mathbf{A}\|_F = \sqrt{\operatorname{tr}(\mathbf{A}^\mathsf{H}\mathbf{A})}$ denotes the Frobenius norm, and $\log$ denotes the principal matrix logarithm.

The LE metric provides another Riemannian structure on HPD manifolds. The LE metric is constructed by mapping HPD matrices to the vector space of Hermitian matrices through the matrix logarithm. The matrix logarithm $\log: \mathscr{P}(N,\mathbb{C}) \rightarrow \mathcal{H}(N,\mathbb{C})$ is a diffeomorphism onto the vector space of Hermitian matrices. By pulling back the standard Frobenius inner product $\langle \mathbf{A}, \mathbf{B}\rangle = \operatorname{tr}(\mathbf{A}^\mathsf{H}\mathbf{B})$ from $\mathcal{H}(N,\mathbb{C})$ to $\mathscr{P}(N, \mathbb{C})$, one obtains the LE metric on the tangent space at point $\mathbf{P}$ \cite{ArsignyVincent2006Lmff},
\begin{equation}
    g_{\text{LE}}({\mathbf{P}})(\mathbf{X},\mathbf{Y}) = \operatorname{tr}\left((d\log_{\mathbf{P}} \mathbf{X})(d\log_{\mathbf{P}} \mathbf{Y})\right),
\end{equation}
where $d\log_\mathbf{P}(\cdot)$ denotes the differential of the matrix logarithm. This construction induces a flat Riemannian geometry for HPD manifolds. Under the LE metric, the geodesic between two HPD matrices $\mathbf{A}$ and $\mathbf{B}$ is given by
\begin{equation}
    \boldsymbol{\gamma}(t) = \exp\left((1-t)\log \mathbf{A} + t \log \mathbf{B} \right), \quad t \in [0,1].
\end{equation}
Accordingly, the induced LE distance is
\begin{equation}\label{def:LEdistance}
    d_{\text{LE}}(\mathbf{A}, \mathbf{B}) = \|\log \mathbf{A} - \log \mathbf{B}\|_F.
\end{equation}

The BW metric originates from optimal transport theory and defines another Riemannian metric for HPD manifolds \cite{BhatiaRajendra2019OtBd}. Interpreting HPD matrices as covariance matrices of zero-mean Gaussian distributions gives the 2-Wasserstein distance between Gaussian measures, that induces a Riemannian metric for  HPD manifolds. This interpretation has made the BW metric increasingly popular in signal processing and machine learning. For any
$\mathbf{P}\in\mathscr P(N,\mathbb C)$
and tangent matrices
$\mathbf{X},\mathbf{Y}\in
T_{\mathbf P}\mathscr P(N,\mathbb C)$,
the BW metric is defined by
\begin{equation}
    g_{\text{BW}}({\mathbf{P}})(\mathbf{X}, \mathbf{Y}) = \frac{1}{2} \operatorname{tr}\left( \mathbf{L}_{\mathbf{P}}(\mathbf{X}) \, \mathbf{Y} \right),
\end{equation}
where $\mathbf{L}_{\mathbf{P}}(\mathbf{X})$ is the unique Hermitian solution to the Lyapunov equation \begin{equation} 
    \mathbf{P} \mathbf{L}_{\mathbf{P}}(\mathbf{X}) + \mathbf{L}_{\mathbf{P}}(\mathbf{X}) \mathbf{P} = 2\mathbf{X}.
\end{equation} 
The corresponding geodesic connecting $\mathbf{A}, \mathbf{B} \in \mathscr{P}(N, \mathbb{C})$ takes the form
\begin{equation}
\begin{aligned}
    \gamma(t)&=\left((1-t)\mathbf{I}+t\mathbf{Q}\right)\mathbf{A}\left((1-t)\mathbf{I}+t\mathbf{Q}\right),\\
    \mathbf{Q}&=\mathbf{A}^{-1/2}\left(\mathbf{A}^{1/2}\mathbf{B}\mathbf{A}^{1/2}\right)^{1/2}\mathbf{A}^{-1/2}.  
\end{aligned}
\end{equation}
The corresponding geodesic induces the BW distance
\begin{equation}\label{def:BWdistance}
    d_{\text{BW}}(\mathbf{A}, \mathbf{B}) = \sqrt{\operatorname{tr}(\mathbf{A}) + \operatorname{tr}(\mathbf{B}) - 2 \operatorname{tr}\left( \left( \mathbf{A}^{1/2} \mathbf{B} \mathbf{A}^{1/2} \right)^{1/2} \right)}.
\end{equation}

Building upon the Riemannian structures, we next address the problem of estimating representative HPD matrices. We introduce the estimation of six representative statistics on  HPD manifolds, namely the means and medians under the AIRM, the LE metric and the BW metric. In particular, the LE mean of HPD matrices $\{\mathbf{R}_1,\mathbf{R}_2,\ldots,\mathbf{R}_{K}\}$ can be calculated by  \cite{ArsignyVincent2006Lmff}
\begin{equation}
    \mathbf{R}_{\text{LE mean}} = \exp\left(\frac{1}{K} \sum\limits_{k=1}^K \log \mathbf{R}_k\right).
\end{equation}
The remaining five statistics do not admit closed forms, and can be obtained by solving optimization problems \cite{OnoYusuke2024TCoR}
\begin{align}
    \mathbf{R}_{\text{mean}} = \underset{\mathbf{R} \in \mathscr{P}(N,\mathbb{C})}{\arg\min}G_{\text{mean}}(\mathbf{R}),
    \end{align}
    and 
    \begin{align}
    \mathbf{R}_{\text{median}} = \underset{\mathbf{R} \in \mathscr{P}(N, \mathbb{C})}{\arg \min} G_{\text{median}}(\mathbf{R}),
\end{align}
respectively. Here, the objective functions are
\begin{align}
    G_{\text{mean}} (\mathbf{R}) &= \frac{1}{K} \sum\limits_{k=1}^K d^2(\mathbf{R}_k, \mathbf{R}),\label{def:RDmeanobj}
    \end{align}
    and
    \begin{align}
    G_{\text{median}} (\mathbf{R}) &= \frac{1}{K} \sum\limits_{k=1}^K d(\mathbf{R}_k, \mathbf{R}).\label{def:RDmedianobj}
\end{align}
where $d(\cdot,\cdot)$ denotes the distance of two HPD matrices, and in the current study, we consider the AIRM \eqref{def:AIRMdistance}, LE \eqref{def:LEdistance} and BW distances \eqref{def:BWdistance}.

The specific algorithm depends on the underlying Riemannian metric and the associated optimization problem. For the AIRM mean, the optimization problem is commonly solved via the Riemannian gradient descent method \cite{OnoYusuke2022Tams}
\begin{equation}
    \mathbf{R}_{t+1} =  \mathbf{R}_t^{\frac{1}{2}} \exp\left(\frac{1}{K}\sum\limits_{k=1}^K\log\left(\mathbf{R}_t^{-\frac{1}{2}} \mathbf{R}_k \mathbf{R}_t^{-\frac{1}{2}}\right)\right)\mathbf{R}_t^{\frac{1}{2}}.
\end{equation}
The BW mean can be calculated by the fixed-point algorithm \cite{BhatiaRajendra2019OtBd}
\begin{equation}
    \mathbf{R}_{t + 1} = \mathbf{R}_t^{-\frac{1}{2}}
    \left(\frac{1}{K}\sum\limits_{k=1}^K\left(\mathbf{R}_t^{\frac{1}{2}} \mathbf{R}_k \mathbf{R}_t^{\frac{1}{2}}\right)^{\frac{1}{2}}\right)^2
    \mathbf{R}_t^{-\frac{1}{2}}.
\end{equation}

For the geometric medians, we will use the Weiszfeld algorithms \cite{FletcherPThomas2009Tgmo}. For the AIRM median, the update is given by
\begin{equation}
    \mathbf{R}_{t + 1} = \mathbf{R}_t^{\frac{1}{2}} \exp\left(\mathbf{Z}_t\right)\mathbf{R}_t^{\frac{1}{2}},
\end{equation}
where
\begin{equation}
\begin{aligned}
    \mathbf{Z}_t &= \left(\sum\limits_{k=1}^K\frac{\log\left(\mathbf{R}_t^{-\frac{1}{2}}\mathbf{R}_k \mathbf{R}_t^{-\frac{1}{2}}\right)}{\left\|
    \log\left(\mathbf{R}_t^{-\frac{1}{2}}\mathbf{R}_k \mathbf{R}_t^{-\frac{1}{2}}\right)\right\|_F}\right)\\
        & \quad \times \left(\sum\limits_{k=1}^K\frac{1}{\left\|
    \log\left(\mathbf{R}_t^{-\frac{1}{2}}\mathbf{R}_k \mathbf{R}_t^{-\frac{1}{2}}\right)
    \right\|_F}\right)^{-1}.
\end{aligned}
\end{equation}
The update of LE median is given by
\begin{equation}
    \mathbf{R}_{t + 1} = \exp\left(\frac{\sum\limits_{k=1}^K w_k^{(t)} \log \mathbf{R}_k}{\sum\limits_{k=1}^K w_k^{(t)}}\right),
\end{equation}
where
\begin{equation}
    w_k^{(t)} = \frac{1}{\|\log \mathbf{R}_t - \log \mathbf{R}_k\|_F}.
\end{equation}
For the BW median, the update is given by
\begin{equation}
    \mathbf{R}_{t + 1} = \mathbf{R}_t^{\frac{1}{2}} \mathbf{Z}_t \mathbf{R}_t^{\frac{1}{2}},
\end{equation}
where
\begin{equation}
    \begin{aligned}
       \mathbf{Z}_t = \left(\sum\limits_{k=1}^K\frac{\left(\mathbf{R}_t^{-\frac{1}{2}} \mathbf{R}_k \mathbf{R}_t^{-\frac{1}{2}}\right)^\frac{1}{2}}{d_{\text{BW}}(\mathbf{R}_t, \mathbf{R}_k)}\right) \left(\sum\limits_{k=1}^K\frac{1}{d_{\text{BW}}(\mathbf{R}_t, \mathbf{R}_k)}\right)^{-1}.
    \end{aligned}
\end{equation}

We remark that several other Riemannian metrics have also been proposed on HPD manifolds (see, e.g., \cite{ThanwerdasYann2022Tgom,ThanwerdasYann2023ORmo}). In this work, we focus on the AIRM, and the LE and BW metrics, as they are among the most widely used metrics in matrix-CFAR detection and statistical learning.

\subsection{Hermitian Positive Definite Neural Network}\label{subsec:HPDNet}

\begin{figure*}[htbp]
\centering
\begin{tikzpicture}
\node at (0,4) {Input};
\draw [ultra thick] (0,3) circle [radius=0.7];
\node at (0,3) {$\mathbf{R}_1$};
\draw [->] (1,3) -- (3,3);
\draw [ultra thick] (4,3) circle [radius=0.7];
\node at (2,3.5) {$\operatorname{BiMap}_{\mathbf{W}_1}$};
\node at (4,3) {$\operatorname{ReEig}$};
\draw [ultra thick] (0,1.5) circle [radius=0.7];
\node at (0,1.5) {$\mathbf{R}_2$};
\draw [->] (1,1.5) -- (3,1.5);
\node at (2,2) {$\operatorname{BiMap}_{\mathbf{W}_2}$};
\draw [ultra thick] (4,1.5) circle [radius=0.7];
\node at (4,1.5) {$\operatorname{ReEig}$};
\draw[dash pattern=on 5pt off 3pt, line width=0.6pt]
    (0,0.5) -- (0,-0.5);
\node at (2,0.5) {$\operatorname{BiMap}_{\mathbf{W}_i}$};
\draw [->] (1,0) -- (3,0);
\draw[dash pattern=on 5pt off 3pt, line width=0.6pt]
    (4,0.5) -- (4,-0.5);
\draw [ultra thick] (0,-1.5) circle [radius=0.7];
\node at (0,-1.5) {$\mathbf{R}_K$};
\draw [->] (1,-1.5) -- (3,-1.5);
\draw [ultra thick] (4,-1.5) circle [radius=0.7];
\node at (2,-1) {$\operatorname{BiMap}_{\mathbf{W}_K}$};
\node at (4,-1.5) {$\operatorname{ReEig}$};
\draw [->] (5,3) -- (7,1.5);
\draw [->] (5,1.5) -- (7,1);
\draw [->] (5,0) -- (7,0.5);
\draw [->] (5,-1.5) -- (7,0);
\draw [ultra thick] (8,0.75) circle [radius=0.7];
\node at (8,0.75) {$\operatorname{Avg}$};
\draw [->] (9,0.75) -- (11,0.75);
\node at (12,0.75) {Output};
\draw [ultra thick] (12,0.75) circle [radius=0.7];
\end{tikzpicture}
\caption{HPDNet structure diagram corresponding to \eqref{eqns:HPDNet_s} with a single layer. The expressions of $\operatorname{BiMap}$ and $\operatorname{ReEig}$ are given in \eqref{eqs:BiMap-ReEig}, where $\mathbf{W}_1,\mathbf{W}_2,\ldots,\mathbf{W}_K$ are trainable parameters. $\operatorname{Avg}$ denotes the arithmetic mean.}
\label{fig:HPDNet_s}
\end{figure*}
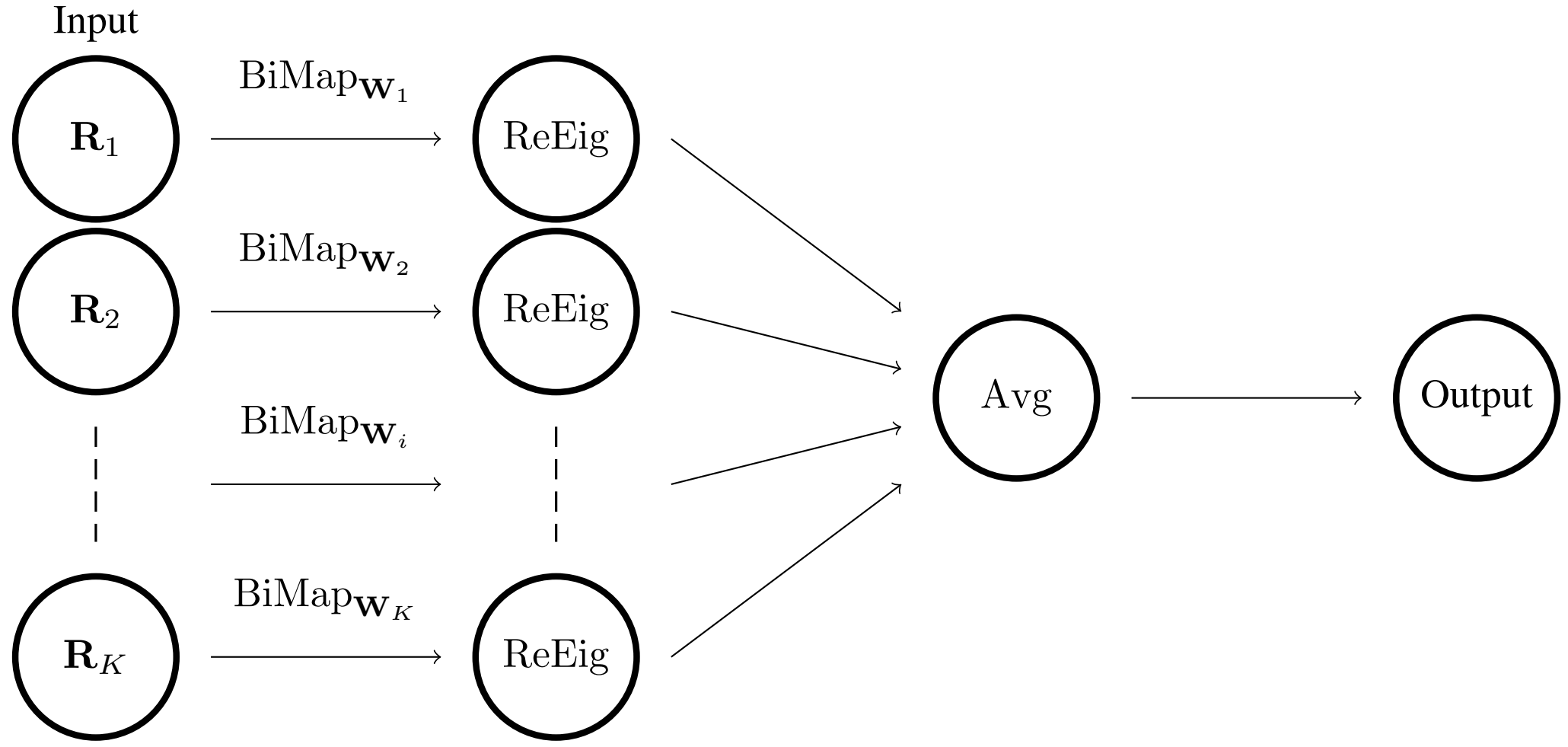

The Riemannian metrics define the geometric structures of HPD manifolds. Building upon this geometric framework, HPDNet performs feature transformations directly on HPD matrices while preserving their Hermitian positive definiteness. Its two fundamental layers, the bilinear mapping (BiMap) layer and the eigenvalue rectification (ReEig) layer, are respectively defined as \cite{BrooksDaniel2019AHPD}
\begin{equation}\label{eqs:BiMap-ReEig}
\begin{aligned}
    \operatorname{BiMap}_{\mathbf{W}}(\mathbf{R}) &= \mathbf{W}\mathbf{R}\mathbf{W}^{\mathsf{H}}, \\
    \operatorname{ReEig}(\mathbf{R};\epsilon) &= \mathbf{U}_{\mathbf{R}} \max(\epsilon \mathbf{I}, \mathbf{\Sigma}_{\mathbf{R}}) \mathbf{U}_{\mathbf{R}}^{\mathsf{H}},
\end{aligned}
\end{equation}
where $\mathbf{R} = \mathbf{U}_{\mathbf{R}}\mathbf{\Sigma}_{\mathbf{R}}\mathbf{U}_\mathbf{R}^{\mathsf{H}}$ denotes the eigenvalue decomposition of an HPD matrix $\mathbf{R}\in  \mathscr{P}(N,\mathbb{C})$ with unitary $\mathbf{U}_{\mathbf{R}}$ and diagonal $\mathbf{\Sigma}_{\mathbf{R}}$, $\mathbf{W} \in \mathbb{C}^{M \times N}$ is the trainable transformation matrix, $M$ denotes the output dimension of the BiMap layer, and $\epsilon$ is a small fixed positive number. In this paper, the HPDNet structure adopted is
\begin{equation}\label{eqns:HPDNet_s}
\begin{aligned}
\mathbf O_k &= \operatorname{ReEig}\left(\operatorname{BiMap}_{\mathbf {W}_k}\left(\mathbf R_{k}\right)\right), \\
&  \qquad k=1,\ldots,K,\\
\operatorname{HPDNet}\left(\mathbf{W},\{\mathbf{R}_k\}_{k=1}^K\right) &= \operatorname{Avg}(\mathbf O_1,\ldots,\mathbf O_K),
\end{aligned}
\end{equation}
where $\operatorname{Avg}$ computes the arithmetic mean of the intermediate matrices $\mathbf O_1,\mathbf O_2,\ldots,\mathbf O_K$. The architecture of HPDNet with a single layer is illustrated in Fig. \ref{fig:HPDNet_s}. Although HPDNet can be constructed with multiple layers, this study focuses exclusively on the approximation capability of its single-layer configuration.

One important property of this architecture is that it preserves the HPD structure throughout the network. Although the output of the $\operatorname{BiMap}$ layer may become rank deficient when $\mathbf{W}$ is not full rank, the subsequent $\operatorname{ReEig}$ layer restores positive definiteness by replacing every eigenvalue smaller than $\epsilon$ with $\epsilon$. Consequently, the output of
\[
\operatorname{ReEig}\left(\operatorname{BiMap}_{\mathbf W}(\mathbf R)\right),\quad \forall \mathbf{R} \in \mathscr{P}(N, \mathbb{C})
\]
is always HPD, allowing the transformation matrices $\mathbf W$ to be optimized as unconstrained in $\mathbb{C}^{M \times N}$.

To integrate the HPDNet into the matrix-CFAR framework, we adopt only its geometry-preserving $\operatorname {BiMap}$ and $\operatorname{ReEig}$ layers. Although existing HPDNet architectures further employ a $\operatorname{LogEig}$ layer to map HPD matrices into Euclidean vectors \cite{BrooksDaniel2019AHPD}, such a mapping is unnecessary in our setting because the objective is to estimate a representative HPD covariance matrix rather than a vector representation. Therefore, the LogEig layer is omitted, allowing the network output to remain on the HPD manifold and to be directly used by the subsequent matrix-CFAR detector, where the geometric objectives are directly used as training loss. Accordingly, the loss functions for estimating the intrinsic mean and median are defined using the Riemannian distances as follows:
\begin{align*}
    \mathcal{L}_{\text{mean}}(\mathbf{W})
    &= \frac{1}{K}\sum_{k=1}^K
    d^2\left(\operatorname{HPDNet}\left(\mathbf{W},\{\mathbf{R}_k\}_{k=1}^K\right) ,\mathbf{R}_k\right),\\
    \mathcal{L}_{\text{median}}(\mathbf{W})
    &= \frac{1}{K}\sum_{k=1}^K
    d\left( \operatorname{HPDNet}\left(\mathbf{W},\{\mathbf{R}_k\}_{k=1}^K\right) ,\mathbf{R}_k\right).
\end{align*}
The model is trained end-to-end using backpropagation. We employ the AdamW optimizer \cite{LoshchilovIlya2019Dwdr}, which decouples weight decay from the gradient update and has demonstrated strong empirical performance for deep neural network optimization.

\section{Detection Performance} \label{sec:performance}
In this section, we evaluate the detection performance of the HPDNet framework in nonhomogeneous clutter environments using both simulated compound-Gaussian clutter and real sea clutter data. 
The performances of the traditional matrix-CFAR and the HPDNet-based approach are analyzed according to their effectiveness and efficiency. All simulations were implemented in PyTorch \cite{PaszkeAdam2019PAis} and conducted on a single GPU.

\subsection{Simulated Clutter Data} \label{sec:result-simulated}
The simulated clutter is modeled as a compound-Gaussian process, which is widely adopted for representing nonhomogeneous clutter due to its ability to capture both local power fluctuations and heavy-tailed amplitude distributions. It therefore provides a realistic benchmark for evaluating matrix-CFAR detectors under nonhomogeneous environments. 

The compound-Gaussian clutter is expressed as
\begin{equation}
\mathbf{c} = \sqrt{\tau} \cdot \mathbf{u}. \label{eqn:compoundGaussian}
\end{equation}
The texture component $\tau$ is assumed to follow a Gamma distribution,
\begin{equation}
\tau \sim \mathrm{Gamma}(\alpha,\beta)
=
\frac{\beta^\alpha}{\Gamma(\alpha)}
\tau^{\alpha-1}e^{-\beta\tau},
\qquad \tau>0,
\end{equation}
where $\alpha>0$ and $\beta>0$ denote the shape and rate parameters, respectively, and $\Gamma(\cdot)$ is the Gamma function. The speckle component $\mathbf{u}$ is modeled as a fast-varying complex Gaussian random vector,
\begin{equation}
\mathbf{u}\sim\mathcal{CN}(\mathbf{0},\mathbf{\Sigma}).
\end{equation}
The speckle covariance $\mathbf{\Sigma}$ is chosen as an exponentially correlated Toeplitz matrix, which is widely used to represent clutter returns from multiple pulses. The $(i,j)$-th entry of $\mathbf{\Sigma}$ is given by
\begin{equation}
\mathbf{\Sigma}_{ij} = \sigma_c^2 r^{|i-j|} e^{\mathrm{j}2\pi f_{d,c} (i-j)} + \sigma_n^2 \delta_{ij},
\end{equation}
where $\sigma_c^2$ is the clutter power, $r \in (0,1)$ is the one-lag correlation coefficient, $f_{d,c}$ is the normalized Doppler frequency of the clutter, $\sigma_n^2$ is the noise power, and $\delta_{ij}$ is the Kronecker delta. The jammer signal is added to randomly selected range cells to evaluate its ability to suppress jamming signals,
\begin{equation}
\mathbf{x}_i^{\text{(jammer)}} = \mathbf{c}_i + \sqrt{\eta \cdot \sigma_c^2} \cdot \mathbf{p}(f_{d,j}), \quad 1\leq i \leq N.
\end{equation}
where $\eta$ is the linear jammer-to-noise ratio, $\sigma_c^2$ is the clutter power, and $\mathbf{p}(f)$ is the Doppler steering vector defined as
\begin{equation}\label{def:dopplersteeringvector}
\mathbf{p}(f) = \frac{1}{\sqrt{N_p}} \left[1, e^{\mathrm{j}2\pi f}, e^{\mathrm{j}4\pi f}, \ldots, e^{\mathrm{j}2\pi (N_p-1)f}\right]^\mathsf{T},
\end{equation}
with $N_p$ being the number of pulses. The signal model is given by
\begin{equation}
\mathbf{x}_{N}^{\text{(target)}} = \mathbf{c}_{N} + \alpha_t \cdot \mathbf{p}(f_{d,t}),
\end{equation}
where $\mathbf{c}_N$ is the clutter vector of the last range cell, $\mathbf{p}(f_{d,t})$ is the target Doppler steering vector defined in \eqref{def:dopplersteeringvector}, and $f_{d,t}$ is the target normalized Doppler frequency. The target amplitude $\alpha_t$ is determined by the signal-to-clutter ratio (SCR): 
\begin{equation}
\alpha_t = \sqrt{\rho \cdot P_c},
\end{equation}
with $\rho$ being the linear SCR and $P_c$ being the estimated clutter power obtained from the median of the first $K - 1$ range cells. By varying $\rho$, we evaluate the detection probability under different SCR conditions. 

\begin{figure}[htbp]
    \centering
    \includegraphics[width=\linewidth]{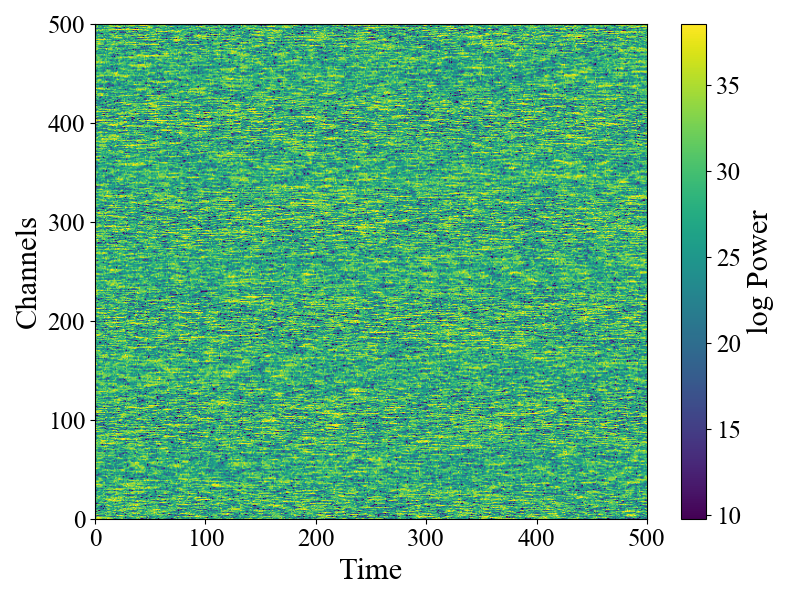}
    \caption{Example of the log power of the compound-Gaussian clutter data.}
    \label{fig:compoundGaussian}
\end{figure}

In the simulations, we set $\sigma_c^2 = 10$ dB, $r = 0.9$, $f_{d,c} = 0.1$, $\eta = 100$, and $\sigma_n^2 = 1$  dB. The compound-Gaussian clutter parameters are set as follows: the texture component $\tau$ follows a Gamma distribution with shape parameter $\alpha = 4$ and rate parameter $\beta = 3$. Two jammers are introduced with a jammer-to-noise ratio of $20$ dB and a normalized Doppler frequency of $f_{d,j} = 0.2$. Fig. \ref{fig:compoundGaussian} shows an example of the compound-Gaussian clutter power with size $500 \times 500$. For each metric, 5000 data blocks of size $5 \times 10$ are used to construct the covariance matrix by \eqref{def:toeplitz}. Thus, the range cell number is $K = 5$ and pulse number is $N_p = 10$. These matrices are fed into the HPDNet as inputs. To determine the matrix-CFAR threshold, 10000 data blocks are used in the Monte Carlo procedure under a constant false alarm rate of $P_{fa} = 0.001$. Then, 2000 data blocks are used to evaluate the detection probability. The target signal injected for detection performance evaluation uses a normalized Doppler $f_{d,t} = 0.1$. The SCR varies from $5$ dB to $30$ dB in $5$ dB increments, where $\rho$ is the corresponding linear SCR value ($\rho = 10^{\text{SCR(dB)}/10}$). 

\begin{figure}[htbp]
    \centering
    \includegraphics[width=\linewidth]{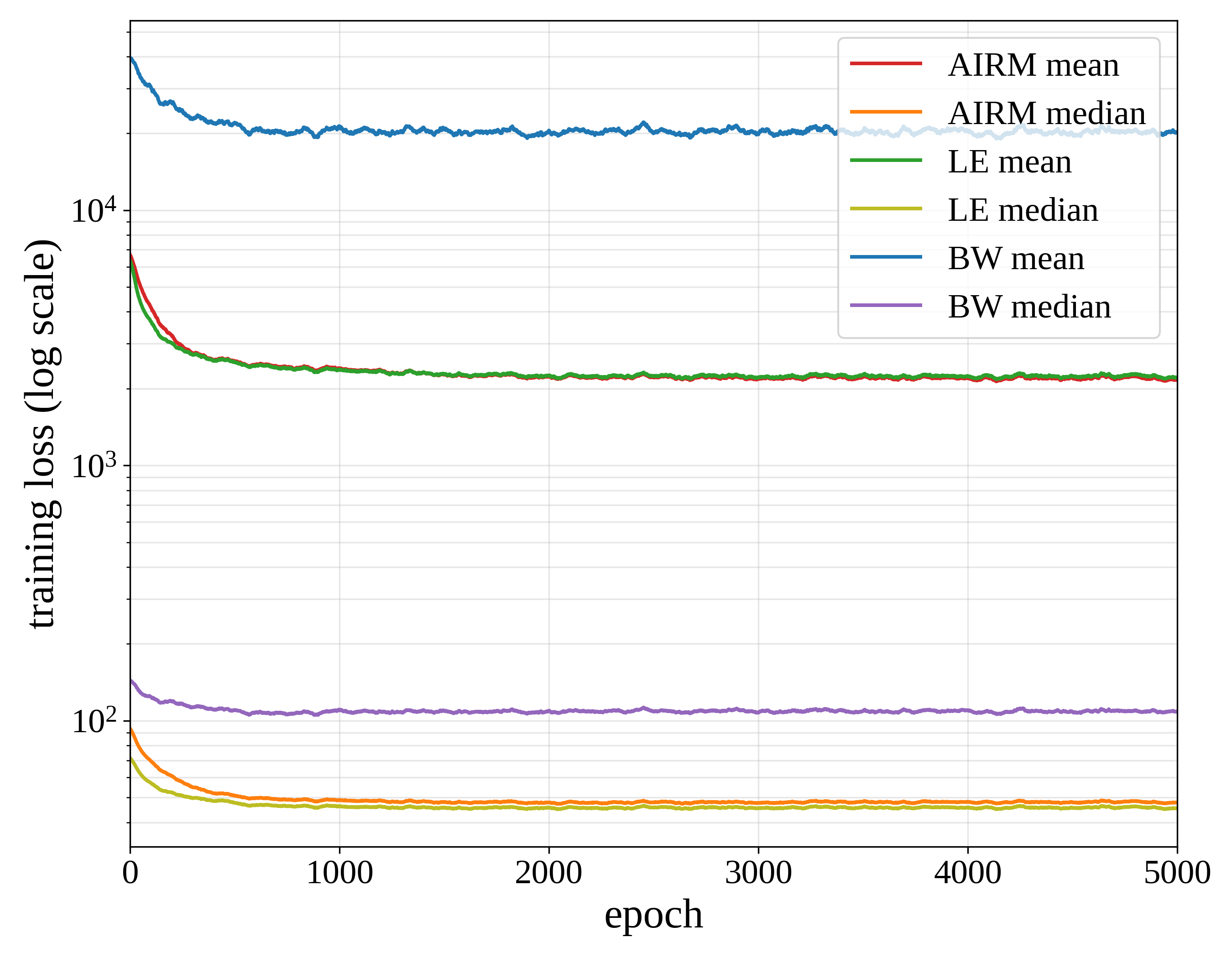}
    \caption{Training loss curves smoothed with a moving window of size 50.}
    \label{fig:train-loss}
\end{figure}

To establish a consistent stopping criterion for comparing the computational cost of different estimation methods, traditional  iterative algorithms terminate when $\|\mathbf{R}_{t+1} - \mathbf{R}_t\|_F < 10^{-6}$. Unlike the traditional iterative processes that adopt a tolerance, we fix 5000 epochs as the termination condition for the HPDNet. This is because the optimized points of loss functions under different metrics are different. Therefore, in practice, we terminate the training once the training loss converges. As shown in  Fig. \ref{fig:train-loss}, the epoch number of 5000 is sufficient for all of the statistics in this paper to satisfy the condition.

\begin{figure*}[htbp]
    \centering
    \subfloat[AIRM (simulated)]{\includegraphics[width = 0.5\textwidth]{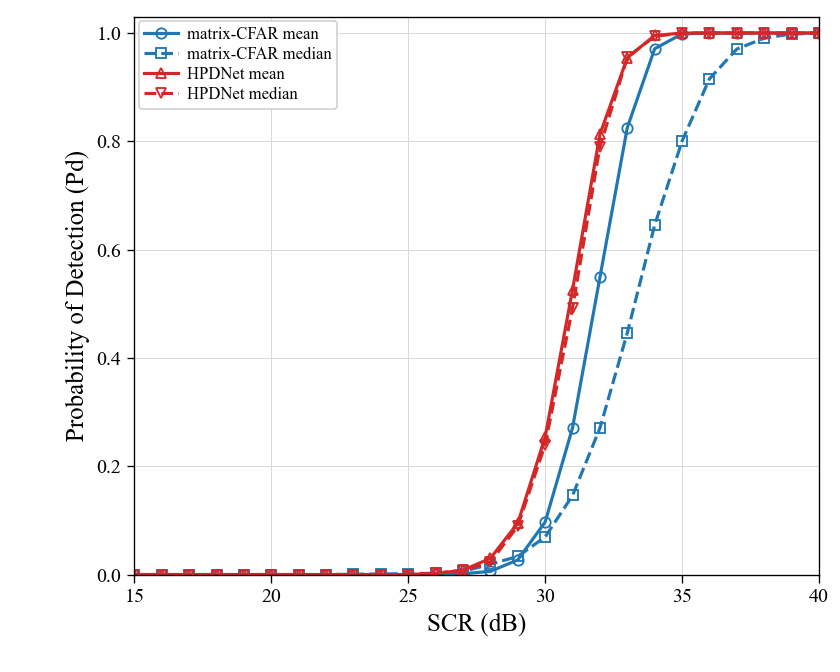}\label{fig:simulation-data-airm}}
    \subfloat[AIRM (real)]{\includegraphics[width = 0.5\textwidth]{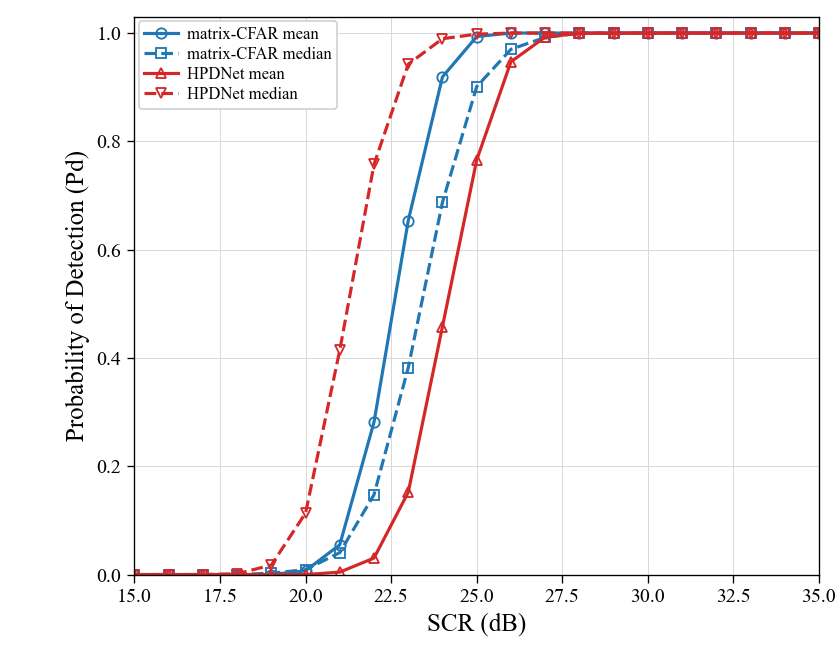}\label{fig:real-data-airm}}\\
    \subfloat[LE (simulated)]{\includegraphics[width = 0.5\textwidth]{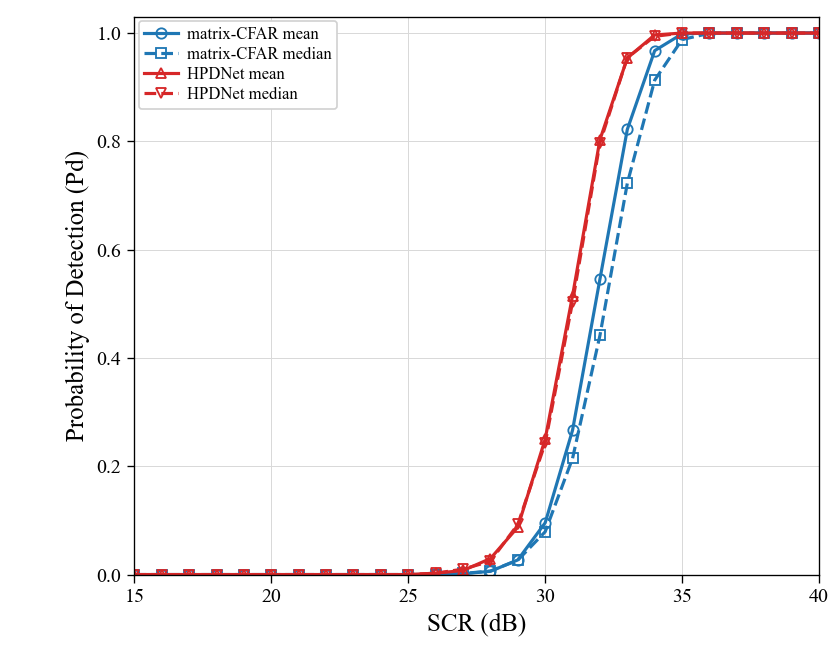}\label{fig:simulation-data-LE}}
    \subfloat[LE (real)]{\includegraphics[width = 0.5\textwidth]{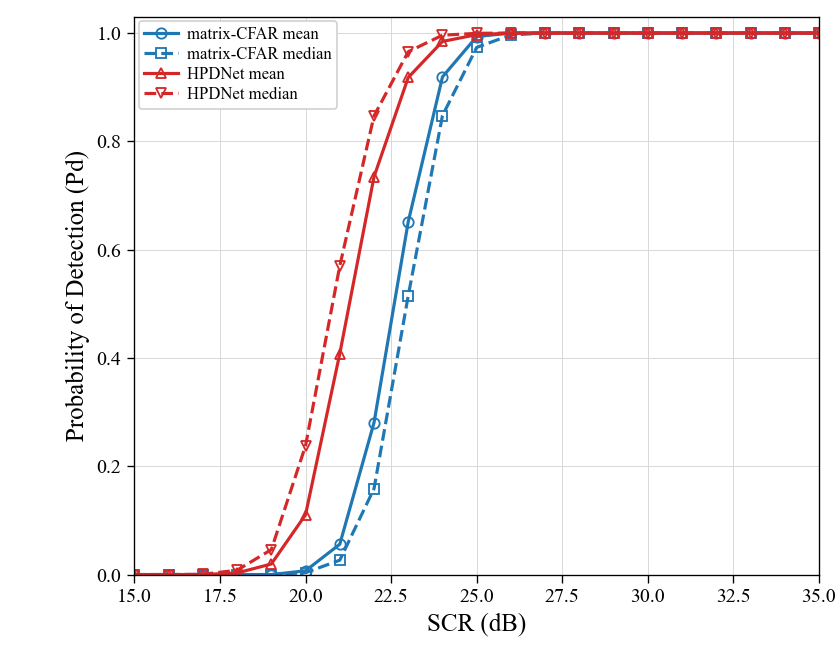}\label{fig:real-data-LE}}\\
    \subfloat[BW (simulated)]{\includegraphics[width = 0.5\textwidth]{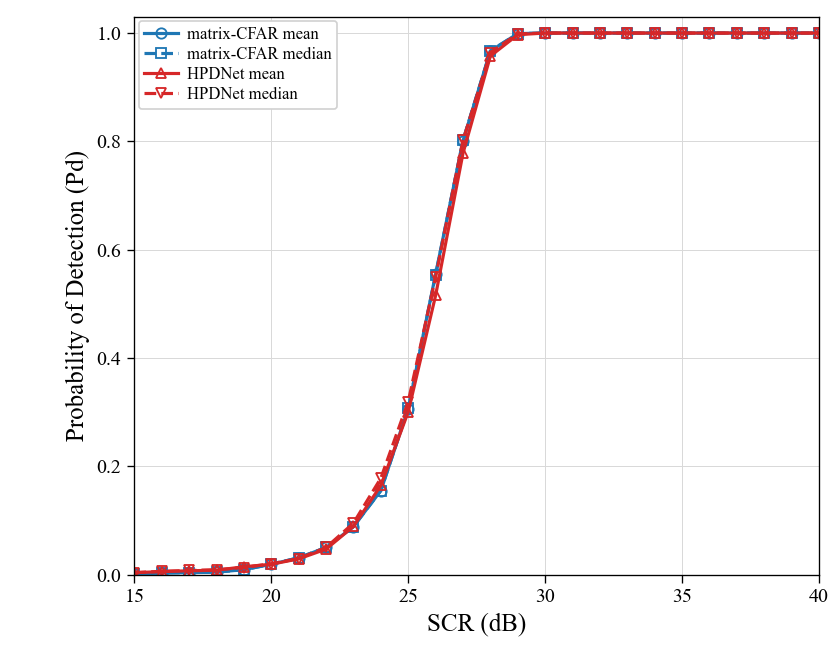}\label{fig:simulation-data-BW}}
    \subfloat[BW (real)]{\includegraphics[width = 0.5\textwidth]{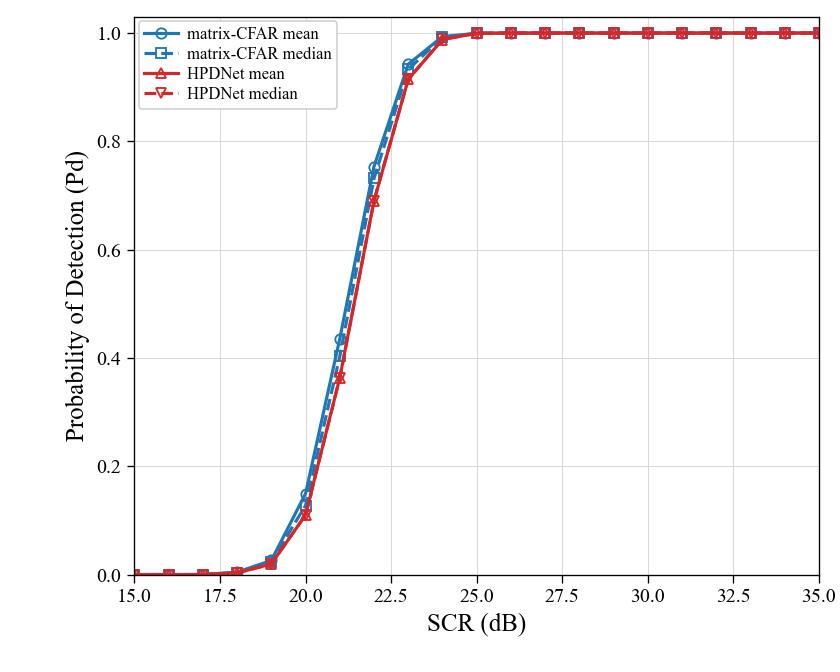}\label{fig:real-data-BW}}
    \caption{The detection result of simulated data (left) and the real data (right).}
\end{figure*}

\begin{figure}[htbp]
    \centering
    \includegraphics[width = 0.5\textwidth]{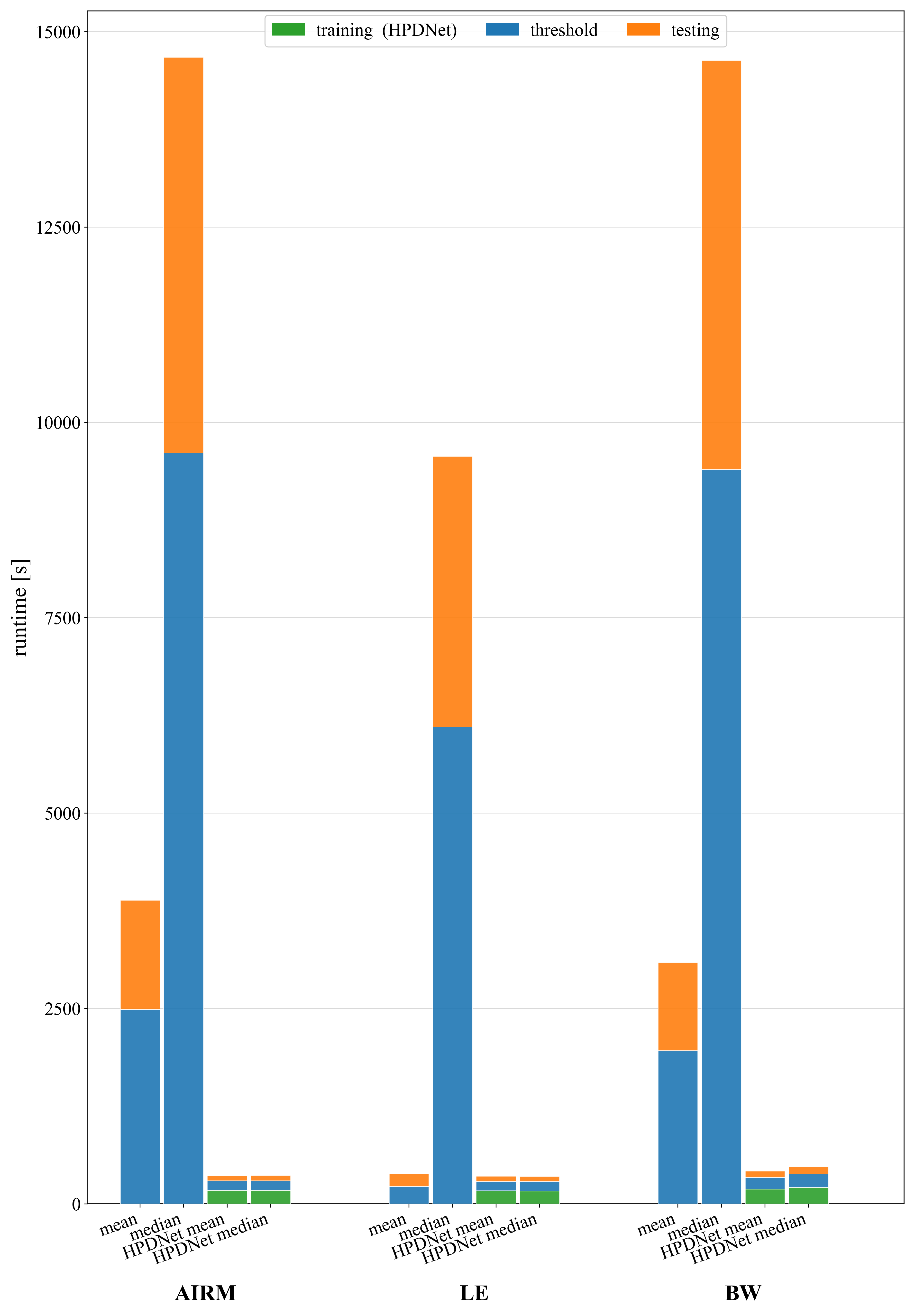}
    \caption{Runtime in simulated data.}
    \label{fig:simulation-data-time}
\end{figure}

The detection results for the compound-Gaussian clutter are shown in Figs. \ref{fig:simulation-data-airm}, \ref{fig:simulation-data-LE}, and \ref{fig:simulation-data-BW}. Under AIRM and LE metric, the detection performance of HPDNet based method is superior than that of the traditional matrix-CFAR method, whereas under BW metric, comparable detection performance is maintained. In addition, the HPDNet significantly reduces detection runtime in all cases, except for the LE mean which admits an analytic formula, as illustrated in Fig. \ref{fig:simulation-data-time}.


\subsection{Real Clutter Data}

\begin{figure}[H]
    \centering
    \includegraphics[width=0.95\linewidth]{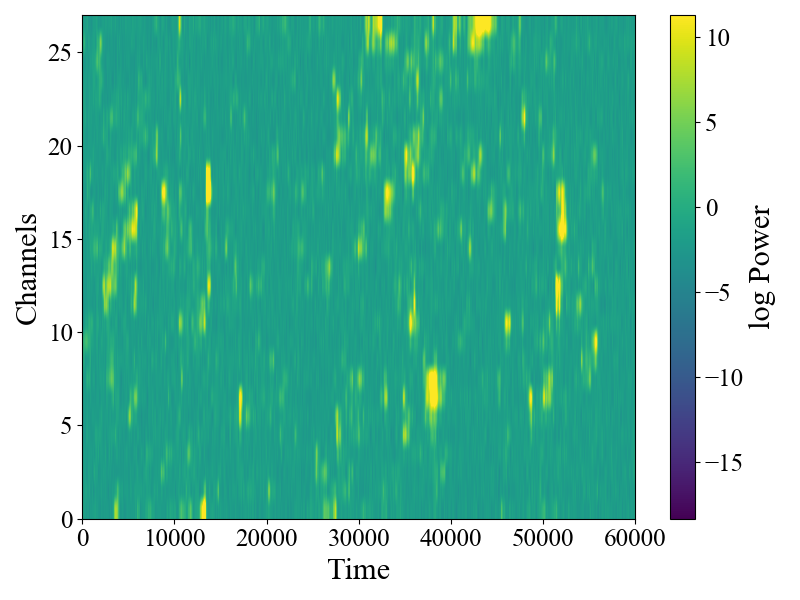}
    \caption{Log power of the real clutter data ``19980205\_192053\_antstep.cdf''.}
    \label{fig:power-19980202-192053}
\end{figure}

We next apply the HPDNet to a real dataset to assess its practical utility. The dataset used here is the IPIX radar sea clutter dataset ``19980205\_192053\_antstep.cdf'' from McMaster University \cite{McMasterIPIX}. Detailed descriptions of the dataset acquisition and experimental configuration can be found in \cite{GiniF2000Performance}. The log power of the data is illustrated in Fig. \ref{fig:power-19980202-192053}. The dataset consists of 27 channels and 60000 time samples. Channels 1 and 27 are excluded from the analysis, and the remaining 25 channels are reshaped into $5 \times 10$ data blocks. As a result, the data are divided into 30000 separate data blocks. As in the simulated-data experiments, 10000 data blocks are used to estimate the threshold through a Monte Carlo procedure with a constant false alarm rate of $P_{fa} = 0.01$. Artificial targets with normalized Doppler frequency $f_{d,t} = 0.2$ are added to the last channel of 2000 data blocks to calculate the detection probability $P_d$. The other parameters are the same as those used in simulated data detection. Note that the HPDNet is trained on simulated compound-Gaussian clutter, as in the simulated-data experiments, and is then directly applied to the IPIX dataset without retraining on real data. This setting is adopted to evaluate the generalization capability of the proposed HPDNet-based matrix-CFAR detector across different clutter environments.

\begin{figure}[htbp]
    \centering
    \includegraphics[width = 0.5\textwidth]{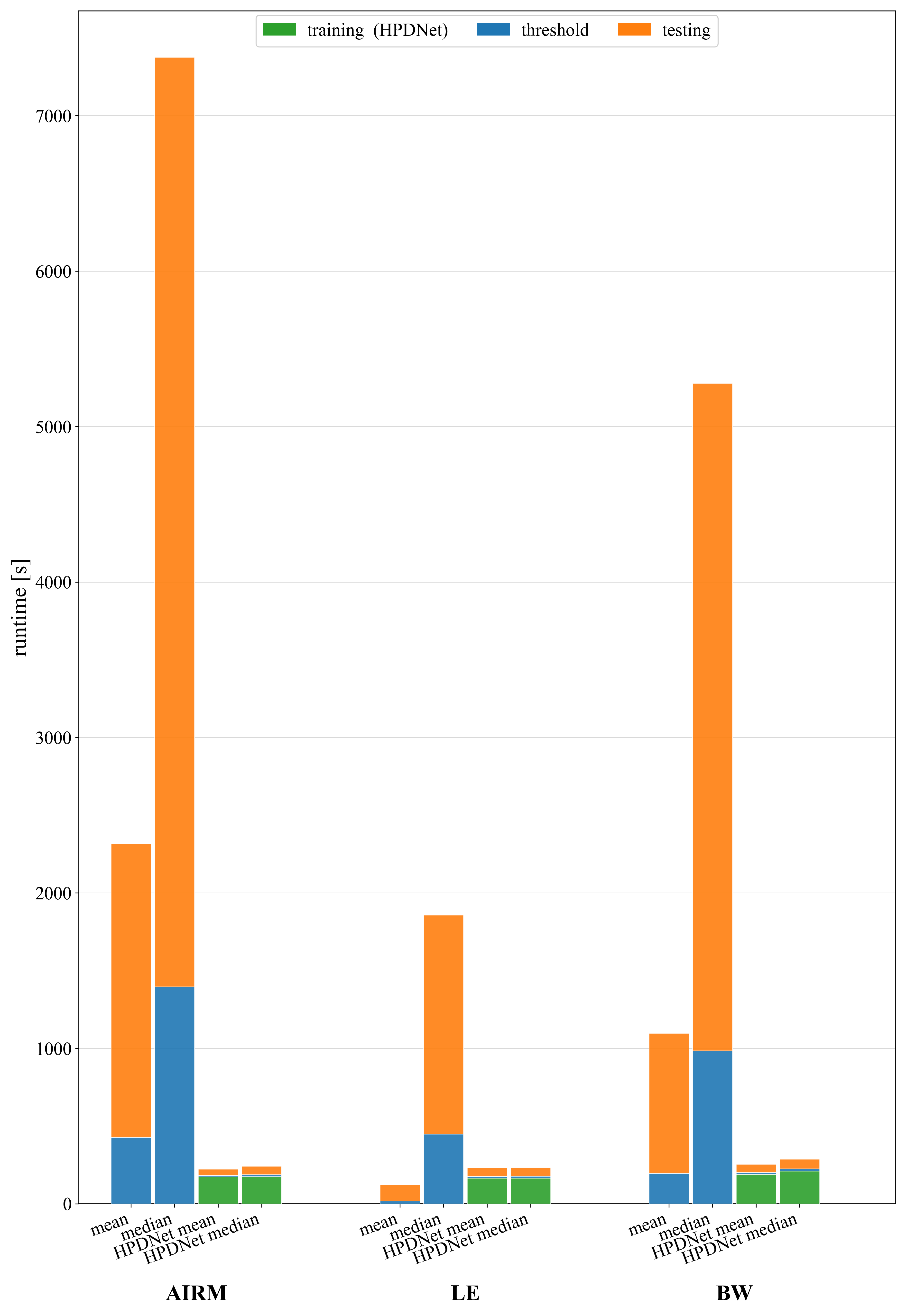}
    \caption{Runtime in real data.}
    \label{fig:real-data-time}
\end{figure}

The detection results for the real sea clutter data are shown in Figs. \ref{fig:real-data-airm}, \ref{fig:real-data-LE} and \ref{fig:real-data-BW}. For AIRM median, the HPDNet achieves superior detection performance while reducing computational cost. However, the case of the AIRM mean does not show a comparable result in detection. Under the LE metric, the HPDNet shows better performance in both the mean and median cases. For BW metric, it maintains comparable performance in detection. Fig. \ref{fig:real-data-time} illustrates that, similar to the simulated case, the testing runtime of the HPDNet is substantially shorter than that of
the traditional iterative methods in all cases, except for the LE mean, which can be computed analytically.

\section{Robustness Analysis}
\label{sec:robustness}

During the practical detection scenarios, outliers might appear to affect the detection when estimating the means or medians \cite{HuaXiaoqiang2021TDWN,OnoYusuke2024TCoR}. 
In this section, we introduce the influence function to evaluate the robustness of HPDNet-based detectors against outliers.

Assume there are $K$ covariance HPD matrices $\{\mathbf{R}_i\}_{i=1}^K$, and  $n$  outliers $\{\mathbf{P}_j\}_{j=1}^n$. Let $\bar{\mathbf{R}}$ be the mean or median of the $K$ HPD matrices without outliers, and $\widehat{\mathbf{R}}$ be the mean or median of the $K + n$ HPD matrices including outliers. The mean or median $\widehat{\mathbf{R}}$ is defined by the following objective functions respectively:
\begin{equation}\label{def:gr0}
    G(\mathbf{R}) := (1 - \varepsilon)\frac{1}{K}\sum\limits_{i=1}^K d^2(\mathbf{R},\mathbf{R}_i) + \varepsilon \frac{1}{n}\sum\limits_{j=1}^n d^2(\mathbf{R},\mathbf{P}_j),
\end{equation}
or
\begin{equation}\label{def:objective-with-outliers}
    G(\mathbf{R}) := (1 - \varepsilon)\frac{1}{K}\sum\limits_{i=1}^K d(\mathbf{R},\mathbf{R}_i) + \varepsilon \frac{1}{n}\sum\limits_{j=1}^n d(\mathbf{R},\mathbf{P}_j).
\end{equation}
Here, $\varepsilon$ is a small positive number.
Treat $\widehat{\mathbf{R}}$ as a perturbation of $\bar{\mathbf{R}}$, we have
\begin{equation}
    \widehat{\mathbf{R}} = \bar{\mathbf{R}} + \varepsilon \mathbf{H}\left(\{\mathbf{R}_i\}_{i=1}^K, \{\mathbf{P}_j\}_{j=1}^n\right) + \mathcal{O}(\varepsilon^2), \label{eqn:perturbation}
\end{equation}
where 
$\mathbf{H}\left(\{\mathbf{R}_i\}_{i=1}^K, \{\mathbf{P}_j\}_{j=1}^n\right)$ is a Hermitian perturbation matrix. Rewriting \eqref{eqn:perturbation} and taking the Frobenius norm yield
\begin{equation}\label{def:hatR-barR}
    \left\|\widehat{\mathbf{R}} - \bar{\mathbf{R}}\right\|_F = \varepsilon \left\|\mathbf{H}(\{\mathbf{R}_i\}_{i=1}^K,\{\mathbf{P}_j\}_{j=1}^n)\right\|_F + \mathcal{O}(\varepsilon^2).
\end{equation}
Then $\left\|\mathbf{H}(\{\mathbf{R}_i\}_{i=1}^K,\{\mathbf{P}_j\}_{j=1}^n)\right\|_F$ can be numerically calculated based on the fact that
\begin{equation}
    \left\|\mathbf{H}(\{\mathbf{R}_i\}_{i=1}^K,\{\mathbf{P}_j\}_{j=1}^n)\right\|_F= \lim_{\varepsilon \rightarrow 0^+} \frac{\left\|\widehat{\mathbf{R}} - \bar{\mathbf{R}}\right\|_F}{\varepsilon}.
\end{equation}
Then the influence function is defined by \cite{OnoYusuke2024TCoR}
\begin{equation}
    f\left(\{\mathbf{R}_i\}_{i=1}^K, \{\mathbf{R}_j\}_{j=1}^n\right) := \frac{\left\|\mathbf{H}\left(\{\mathbf{R}_i\}_{i=1}^K, \{\mathbf{P}_j\}_{j=1}^n\right)\right\|_F}{\|\bar{\mathbf{R}}\|_F}.
\end{equation}
Note that an analytic method for calculating $\mathbf{H}(\{\mathbf{R}_i\}_{i=1}^K,\{\mathbf{P}_j\}_{j=1}^n)$ and hance the influence function can be found in \cite{OnoYusuke2022Tams}.

When outliers are introduced, the number of inputs is increased in the HPDNet, and hence two HPDNets have to be trained, one for determining $\bar{\mathbf{R}}$, and the other for  $\widehat{\mathbf{R}}$. Let $\mathbf{W}_1,\ldots,\mathbf{W}_K$ be  the trained weights  in the first HPDNet with $K$ number of inputs that determines  $\bar{\mathbf{R}}$. In the second HPDNet with $K+n$ inputs for $\widehat{\mathbf{R}}$, the first $K$ weights are fixed as the ones trained from the first $K$ input HPDNet, and we only train the remaining weights $\mathbf{W}_{K+1},\ldots,\mathbf{W}_{K+n}$ using the objective functions \eqref{def:gr0} or \eqref{def:objective-with-outliers} as the loss function. Consequently, $\widehat{\mathbf{R}}|_{\varepsilon =0}=\bar{\mathbf{R}}$. In other words,  in the absence of outliers, $\widehat{\mathbf{R}}$ and $\bar{\mathbf{R}}$ should coincide when computed from the same set of HPD matrices $\{\mathbf{R}_i\}_{i=1}^K$.


In the experiments, we adopt the compound-Gaussian distribution   \eqref{eqn:compoundGaussian} to obtain the influence functions. The number of range cells is set as $K = 50$, and the number of pulses  $N_p = 5$. The outlier-clutter-ratio is fixed at 40 dB. The other parameters of clutter, such as normalized Doppler frequency for clutter, number of jammers, etc., are the same as introduced in Section \ref{sec:result-simulated}. The number of outliers  $n$  varies from 1 to 40.
 
\begin{figure}[htbp]
    \centering
    \subfloat[AIRM]{\includegraphics[width = 0.4\textwidth]{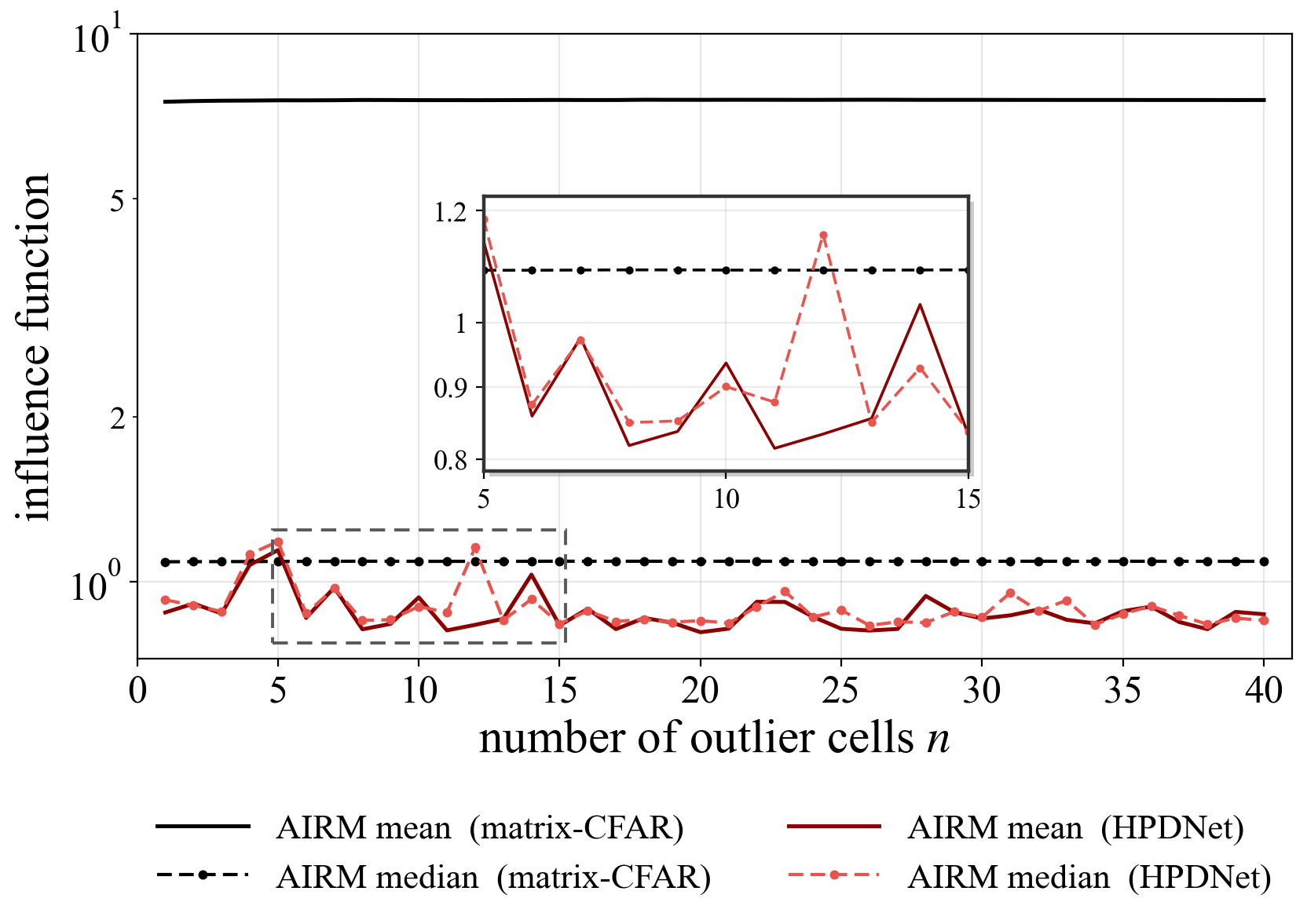}\label{fig:robustness-AIRM}}\\
    \subfloat[LE]{\includegraphics[width = 0.4\textwidth]{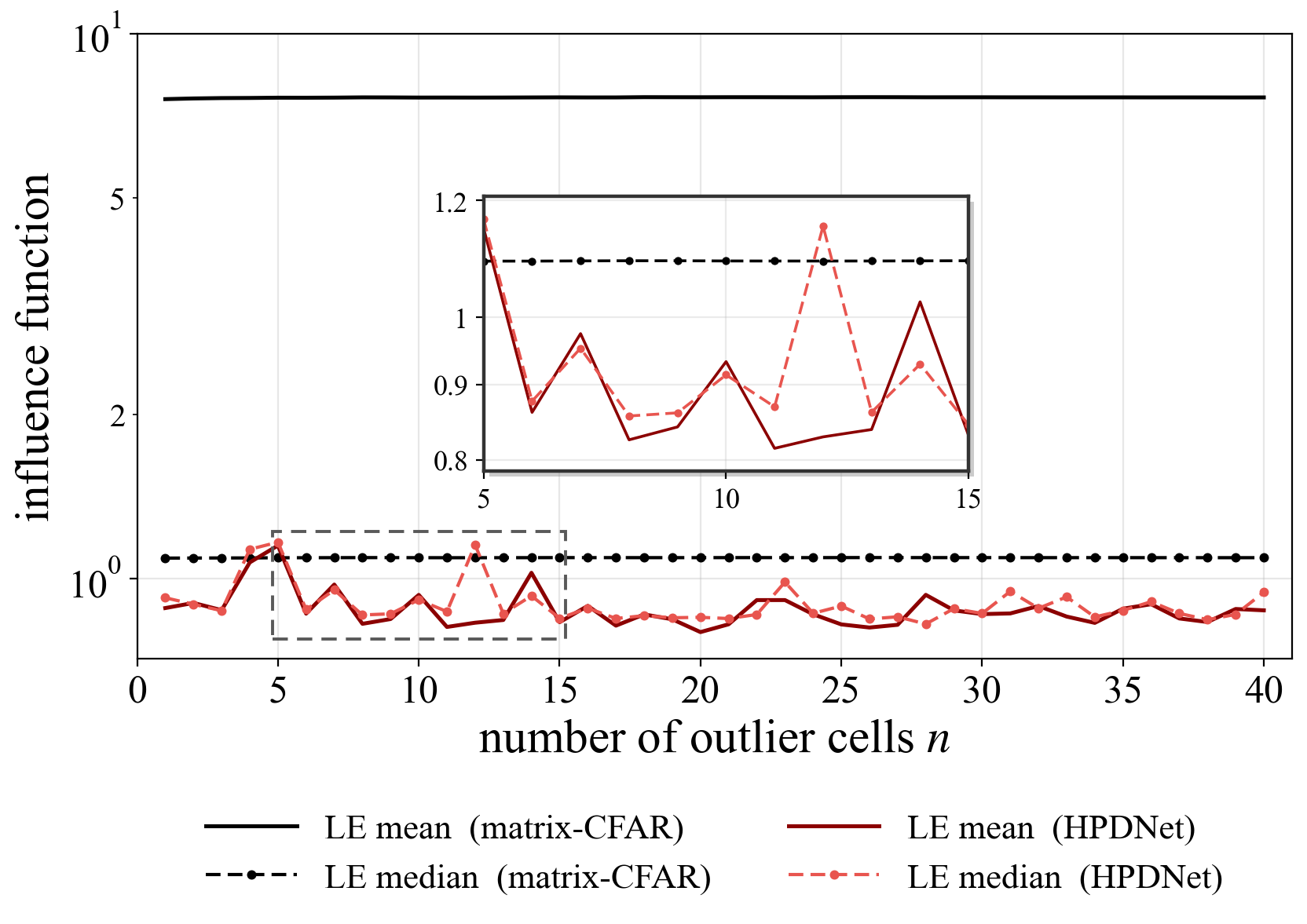}\label{fig:robustness-LE}}\\
    \subfloat[BW]{\includegraphics[width = 0.4\textwidth]{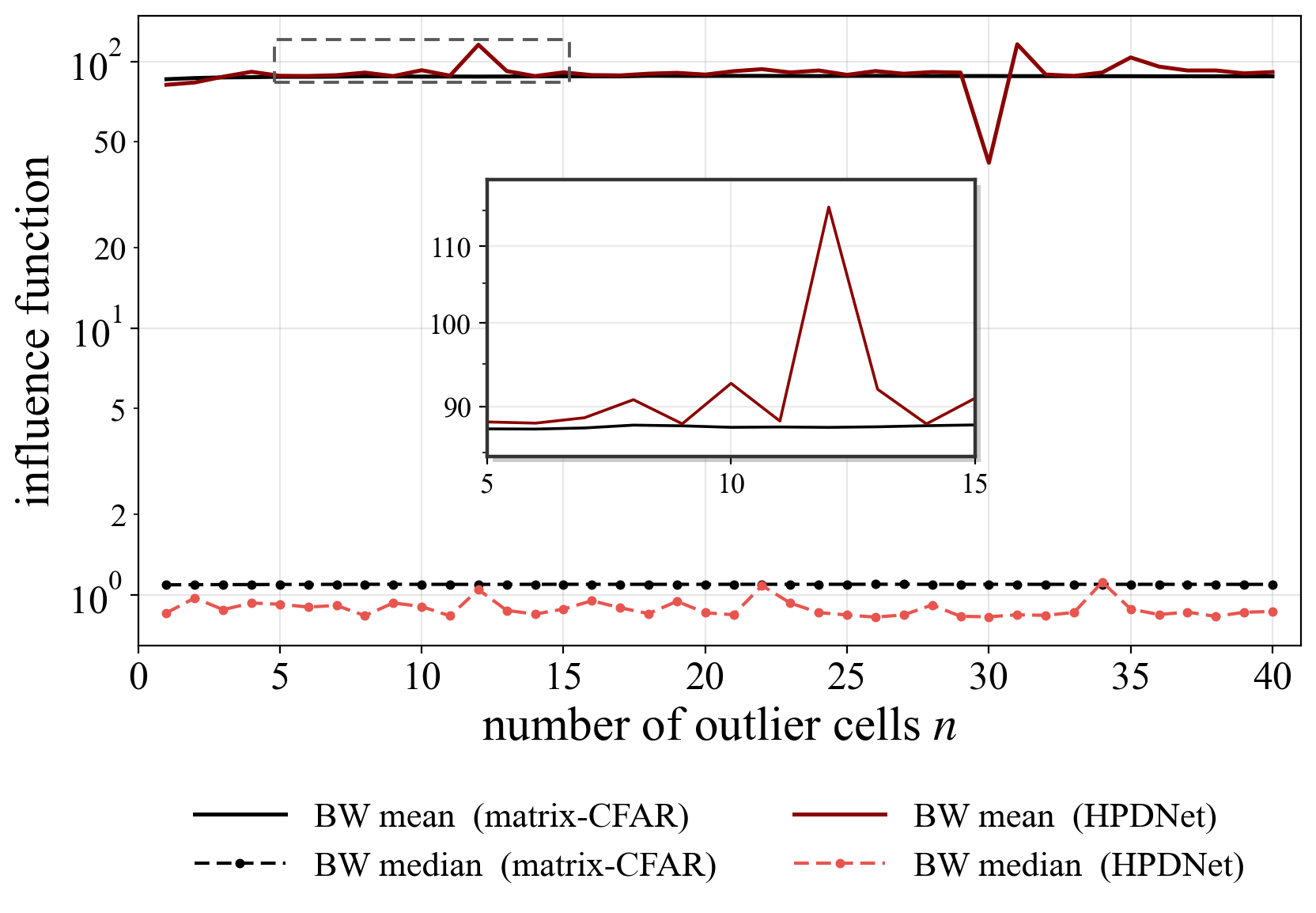}\label{fig:robustness-BW}}\\
    \caption{Influence functions in traditional and HPDNet-based estimations.}
    \label{fig:robustness}
\end{figure}

As shown in Fig. \ref{fig:robustness}, the influence functions of the HPDNet-based estimators exhibit behavior comparable to that of traditional estimators across all three Riemannian metrics, despite some minor differences, particularly in the mean estimators. These results indicate that the proposed HPDNet framework does not compromise robustness against outliers, while maintaining comparable detection performance and significantly reducing computational costs.

\section{Conclusion} \label{sec:conclusion}


We have proposed an HPDNet-based framework for matrix-CFAR geometric signal detection in nonhomogeneous clutter, in which the iterative estimation of geometric means and medians on HPD manifolds is replaced by a learned end-to-end mapping. This approach eliminates the need for iterative optimization at each test instance, thereby enabling rapid inference once the network has been trained.

 Experimental results demonstrate effectiveness of  HPDNet across all three Riemannian metrics considered in this study. For simulated compound-Gaussian clutter, HPDNet improves detection performance under the AIRM and the LE metric while achieving performance comparable to that of conventional detectors under the BW metric. For real IPIX sea clutter data, the HPDNet continues to improve detection performance compared to the classical MIG detectors under the AIRM median and both the LE mean and median. Although no further improvement is observed under the BW metric, HPDNet maintains performance comparable to classical BW detectors without degradation. 
 
 In terms of computational efficiency, HPDNet substantially reduces computational costs compared with the corresponding iterative algorithms used in MIG detectors, with the exception of the LE mean, which admits an analytical expression. Moreover, since HPDNet is trained solely on simulated clutter and can be applied directly to real sea clutter without retraining, these results demonstrate the generalization capability of the learned mapping. Finally, the robustness analysis confirms that HPDNet preserves the robustness of conventional geometric estimators against outliers.

These findings suggest that learning-based geometric processing on HPD manifolds offers a favorable trade-off between detection performance and computational efficiency, making it a promising approach for real-time signal detection. Although this study focuses on the AIRM, and the LE and BW metrics as representative choices, the underlying idea of replacing iterative optimization with a structure-preserving neural mapping is not limited to these three metrics and could be extended to other geometric measures on HPD manifolds \cite{ThanwerdasYann2022Tgom,ThanwerdasYann2023ORmo}. 

Another promising direction is to explore alternative Riemannian neural network architectures. Several architectures have been developed for SPD manifolds, including RResNet \cite{HeKaiming2016DRLf,RiemannianKatsman2023} and SPD Transformer \cite{VaswaniAshish2017AIAY,StructureMathieu2024}, with applications to electroencephalogram (EEG) analysis, facial expression recognition, etc. Investigating how these architectures can be integrated into the matrix-CFAR framework may provide further opportunities to improve detection performance, computational efficiency, and robustness in practical signal detection applications.
\\

{\bf Acknowledgments.}  LP is partially supported by JSPS KAKENHI (JP24K06852) and JST CREST (JPMJCR24Q5).

\bibliographystyle{IEEEtran}

\bibliography{mybibfile}

\end{document}